\documentclass[conference]{IEEEtran}
\IEEEoverridecommandlockouts
\usepackage{graphicx}
\usepackage{float}
\usepackage{xcolor}
\usepackage{csquotes}
\usepackage{booktabs} 
\usepackage{caption} 
\usepackage{amsmath}
\usepackage{amssymb}
\usepackage{array} 
\usepackage{subcaption} 
\usepackage[hidelinks]{hyperref}
\usepackage{algorithm}
\usepackage{algorithmic}
\usepackage[switch]{lineno}
\usepackage{svg}
\svgpath{{../svgs/}}
\usepackage[english]{babel}
\usepackage[
backend=biber,
style=ieee,
sorting=ynt
]{biblatex}
\renewcommand{\thetable}{\arabic{table}} 

\newcommand{\curly}{\mathrel{\leadsto}}
\newcommand{\dimS}{27}
\newcommand{\dimA}{6}
\newcommand{\realR}{\mathbb{R}}

\newcommand{\fiotwo}{\textnormal{FiO}_{2}}
\newcommand{\pip}{\textnormal{P}_{\textnormal{insp}}}
\newcommand{\ti}{\textnormal{T}_{\textnormal{insp}}}
\newcommand{\rr}{\textnormal{RR}}
\newcommand{\peep}{\textnormal{PEEP}}
\newcommand{\tidalvolume}{\textnormal{V}_{\textnormal{T}}}

\newcommand{\paotwo}{\textnormal{PaO}_{\textnormal{2}}}
\newcommand{\spotwo}{\textnormal{SpO}_{\textnormal{2}}}
\newcommand{\cotwo}{\textnormal{CO}_{\textnormal{2}}}
\newcommand{\etcotwo}{\textnormal{etCO}_{\textnormal{2}}}
\newcommand{\hr}{\textnormal{HR}}
\newcommand{\ieratio}{\textnormal{I:E}}

\newcommand{\lineardropoffreward}{R_\textnormal{m}}
\newcommand{\rewardin}{R_\textnormal{in}}
\newcommand{\rewardout}{R_\textnormal{out}}
\newcommand{\lowerb}{\textnormal{LB}}
\newcommand{\upperb}{\textnormal{UB}}

\newcommand{\ph}{\textnormal{pH}}
\newcommand{\pplat}{\textnormal{P}_{\textnormal{plat}}}

\newcommand{\pao}{\textnormal{PaO}_{\textnormal{2}}}
\newcommand{\spo}{\textnormal{SpO}_{\textnormal{2}}}

\newcommand{\cmh}{\textnormal{cmH}_{2} \textnormal{O}}
\newcommand{\mmHg}{\textnormal{mmHg}}
\newcommand{\breathspm}{\textnormal{bpm}}

\newcommand{\va}{\vec{a}}
\newcommand{\vs}{\vec{s}}
\newcommand{\vz}{\vec{z}}
\newcommand{\vsp}{\vec{s'}}
\newcommand{\Tsas}{T(\vsp \mid \vs, \va)}
\newcommand{\Rs}{R(\vs)}
\newcommand{\Ra}{R(\va)}
\newcommand{\Rsa}{R(\vs,\va)}

\newcommand{\dyn}{\textnormal{Dyn}}
\newcommand{\dynlatent}{\textnormal{Dyn}_{\textnormal{L}}}
\newcommand{\enc}{\textnormal{Enc}}
\newcommand{\dec}{\textnormal{Dec}}

\newcommand{\numpat}{100}
\newcommand{\numpathalf}{50}
\newcommand{\Kusedsmpc}{32}
\newcommand{\Hused}{4}
\newcommand{\deltat}{30}

\newcommand{\ardsnet}{ARDSnet }

\renewcommand{\vec}[1]{\mathbf{#1}}

\def\BibTeX{{\rm B\kern-.05em{\sc i\kern-.025em b}\kern-.08em
    T\kern-.1667em\lower.7ex\hbox{E}\kern-.125emX}}
\begin{document}

\title{Optimal Control of Mechanical Ventilators with Learned Respiratory Dynamics}


\author{\IEEEauthorblockN{Isaac R. Ward$^{*,\,\dag}$}\thanks{$^*$Corresponding author.}
\IEEEauthorblockA{
irward@stanford.edu}
\and
\IEEEauthorblockN{Dylan M. Asmar$^{\dag}$\thanks{$^\dag$Stanford Intelligent Systems Laboratory, Department of Aeronautics and Astronautics, Stanford University, Stanford, USA.}}
\IEEEauthorblockA{
asmar@stanford.edu}
\and
\IEEEauthorblockN{Mansur Arief$^{\,\dag}$}
\IEEEauthorblockA{
ariefm@stanford.edu}
\and
\IEEEauthorblockN{Jana Krystofova Mike$^{\ddag}$\thanks{$^\ddag$Department of Pediatrics, University of California San Francisco, UCSF Weill Institute for Neurosciences, San Francisco, USA.}}
\IEEEauthorblockA{
jana.mike@ucsf.edu}
\and
\IEEEauthorblockN{Mykel J. Kochenderfer$^{\dag}$}
\IEEEauthorblockA{
mykel@stanford.edu}
}

\maketitle

\begin{abstract}
Deciding on appropriate mechanical ventilator management strategies significantly impacts the health outcomes for patients with respiratory diseases. Acute Respiratory Distress Syndrome (ARDS) is one such disease that requires careful ventilator operation to be effectively treated. In this work, we frame the management of ventilators for patients with ARDS as a sequential decision making problem using the Markov decision process framework. We implement and compare controllers based on clinical guidelines contained in the \ardsnet protocol, optimal control theory, and learned latent dynamics represented as neural networks. The Pulse Physiology Engine's respiratory dynamics simulator is used to establish a repeatable benchmark, gather simulated data, and quantitatively compare these controllers. We score performance in terms of measured improvement in established ARDS health markers (pertaining to improved respiratory rate, oxygenation, and vital signs). Our results demonstrate that techniques leveraging neural networks and optimal control can automatically discover effective ventilation management strategies \textit{without} access to explicit ventilator management procedures or guidelines (such as those defined in the \ardsnet protocol).
\end{abstract}


\section{Introduction}


Acute Respiratory Distress Syndrome (ARDS) is a clinical syndrome of acute hypoxemic respiratory failure due to lung inflammation (not caused by cardiogenic pulmonary edema), and requires emergency care to prevent further complications or death \cite{matthay2024ardsdef}. 


A common approach to treating ARDS involves mechanical ventilation \cite{matthay2019acute}. Mechanical ventilators (herein `ventilators') are machines that artificially support a patient's breathing by moving air into and out of their lungs according to a set of control inputs that define the pressure, time, fraction of inspired oxygen, etc. of the desired respiratory cycle \cite{slutsky2000mechanical}.


{Our specific focus in this work} is to illustrate that methods from the field of optimal {control}---supplemented with learning-based techniques---are capable of optimally and automatically choosing a ventilator's inputs to effectively manage the condition of a patient with ARDS in real time. {The contributions of this work are as follows}:

\begin{enumerate}
    \item A repeatable ventilator management task (formulated as a Markov Decision Process and built on top of the Pulse Physiology Engine \cite{bray2019pulse}) for assessing the performance of ventilator management strategies.
    
    \item Comparisons of optimal control policies, learning-based policies, and established ventilator management strategies (e.g. the \ardsnet protocol) on this task. 

    \item Evidence demonstrating that neural networks and optimal control can automatically discover effective ventilation management strategies \textit{without} access to explicit ventilation procedures, recommendations, or guidelines. 
\end{enumerate}

A key baseline in this work is the \textit{\ardsnet protocol}, which is a strategy for setting ventilator actions to manage ARDS effectively \cite{grasso2007ardsnet}. It uses a rules-based decision-making framework, calculating control inputs for a ventilator using mathematical formulas and lookup tables based on clinical trial data from the NIH and NHLBI. Other ventilator management protocols also exist, emphasizing specific outcomes such as protective ventilation (e.g. to avoid ventilator-induced lung injury \cite{zou2024ventstratsprotect}).

In addition to traditional protocols, methods for fully automating ventilator controls have been explored. By automating ARDS management, these techniques may offer more constant, fine-grained, and responsive care to patients with ARDS. Approaches include adaptive fuzzy proportional and integral controllers \cite{naskar2023fuzzypi}, the partially observable Markov decision process (POMDP) framework \cite{li2019optimizing,kreutz2005pomdp}, and reinforcement learning \cite{peine2021development}. Our approach builds on these by integrating optimal control theory with learning-based techniques, enabling the discovery of effective ventilator strategies without explicit guidelines.




\section{Modeling Patients with Respiratory Disease}
\label{sec:pulse}

We use the open source Pulse Physiology Engine (or `Pulse' for short) to simulate the respiratory dynamics of ARDS patients in this work. Pulse provides a suite of physiology models and includes a respiratory system simulator capable of simulating ARDS with varying configurations and severities. ARDS (and other respiratory diseases) are simulated in Pulse with a circuit model of the human respiratory system that accounts for the compliances in the chest wall and lungs  \cite{bray2019pulse}.

The pressure-volume relationships that Pulse specifies \cite{harris2005pvcurve,venegas1998pvcurve} have been well-studied, and the engine's modeling has shown empirical agreement with measurements pertaining to both 1) the mechanical behavior of the lungs and 2) patient vitals, particularly when simulating moderate severity ARDS \cite{bray2019pulse}. 


\begin{figure*}[h] 
    \centering
    \includegraphics[width=0.65\textwidth]{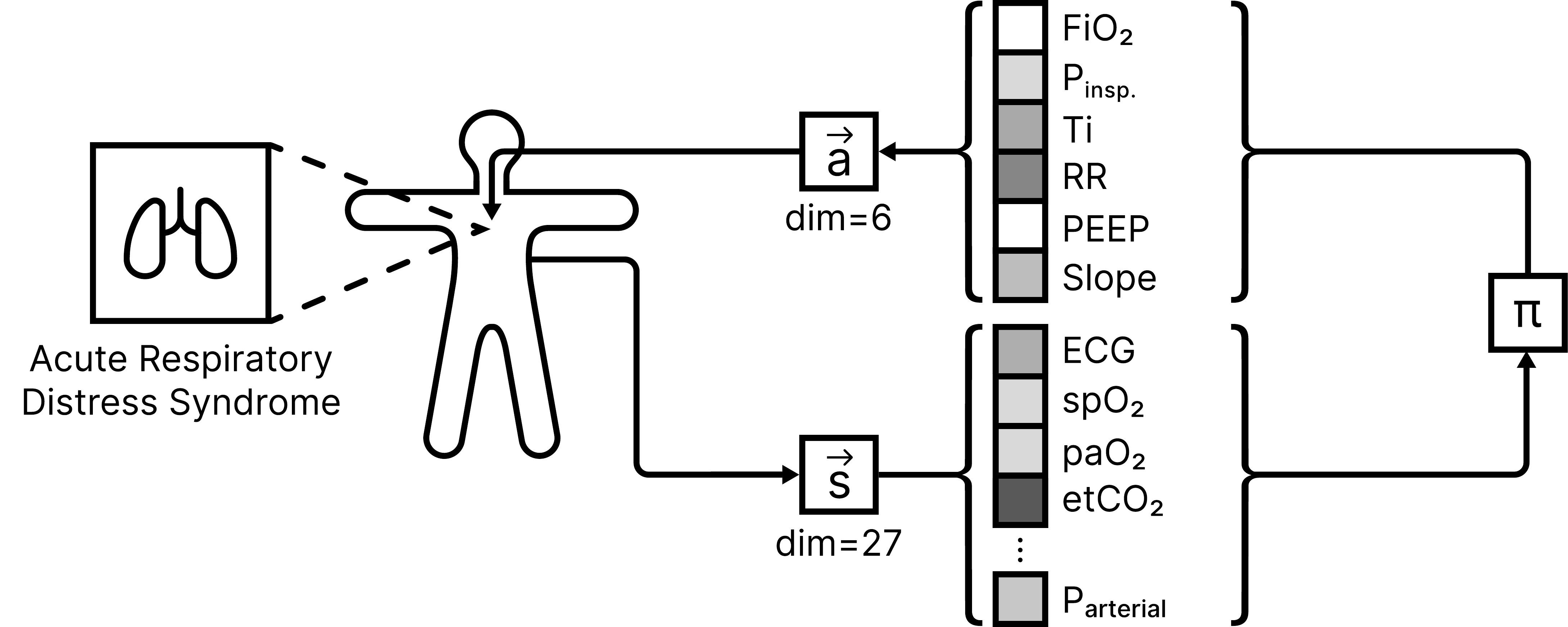}

    \caption{An overview of this work. A policy $\pi$ considers an ARDS patient's health state $\vs \in \realR^{\dimS}$ and determines a ventilator action $\va \in \realR^{\dimA}$ through some decision-making algorithm. This process repeats until $48$ hours have been simulated (or $96$ sequential ventilator actions spaced at $\Delta t = \deltat$ minutes). The algorithms in this work (generally) operate by maximizing some reward $\Rsa$, where a high-reward is associated with desirable outcomes for ARDS patients. Desirable outcomes refer to blood chemistry, heart rate, breathing cycle measurements, etc. being in acceptable ranges for that patient (see Section~\ref{sec:reward}). A full dictionary of all variables is made available in Table~\ref{tab:variables}. The methods by which actions are computed are visualized in Figure~\ref{fig:algorithms}. The Pulse Physiology Engine is used to advance the simulation of the patient's health state (though learned approximations are used to estimate the patient's respiratory dynamics in some policies). } 
    
    \label{fig:overview}
\end{figure*}

\section{Preliminaries \& Problem Formulation}
\label{sec:problem}


We describe the problem using the notation of Markov Decision Processes (MDPs), and formulate it as a $5$-tuple $\langle S, A, T, R, \gamma \rangle$, where:
\begin{itemize}
    \item $S$ is the set of states where $\vec{s}\in S \subseteq \realR^{\dimS} $ describes the patient's health (respiratory, blood gases, vital signs, etc.).
    
    \item $A$ is the set of actions where $\vec{a}\in A \subseteq \realR^{\dimA} $ describes a ventilator action. Specifically, the $\dimA$ numbers comprising any action vector $\vec{a}$ are the ventilator's fraction of inspired oxygen ($\fiotwo$), inspiratory pressure ($\pip$), inspiratory time ($\ti$), respiratory rate ($\rr$), positive end-expiratory pressure ($\peep$), and slope.

    \item $\Tsas$ is the probability of transitioning from state $\vs$ to $\vsp$ given an action $\va$. In other words, $\Tsas$ describes the patient's future health given their current health and ventilator actions. 

    \item $\Rsa$ describes the total reward associated with a patient's health state $\vs$ and the ventilator input $\va$ at a given time. $\Rsa$ in this work is designed to reward recovering patients and protective ventilator strategies, and is defined explicitly in Section~\ref{sec:reward}.

    \item $\gamma$ is the discount factor and describes how we weight future rewards against immediate rewards.
\end{itemize}


Our task is to determine an optimal ventilator policy $\pi$ for ARDS patients (see Figure~\ref{fig:overview}). In other words, at time $t$, we want to find a strategy that determines an optimal ventilator action $\va^*_t$ by considering the patient's current health state $\vs_t$ such that the patient's health improves throughout ventilation. 

Our benchmark simulates $\numpat$ patients. The {demographics} are $\numpathalf$ male and $\numpathalf$ female, with ages distributed uniformly over the range $18$--$65$, each presenting with mild-to-severe ARDS. Each patient's health must be optimized throughout $48$ hours of ventilation, with our policies $\pi$ determining a sequence of $96$ ventilator actions (a rate of one ventilator update every $\Delta t=\deltat$ minutes).

\subsection{Rewards}
\label{sec:reward}

We decompose the reward function into $\Rsa=\Rs+\Ra$, where $\Rs$ and $\Ra$ evaluate the optimality of 1) patient health states and 2) ventilator actions respectively. 

We {evaluate the patient's health state} by generating a positive reward $\rewardin$ for each health marker $x$ if $x$ falls within the bounds $[\lowerb_x, \upperb_x]$ (see Equation~\ref{eq:rcomponent}). As $x$ falls further outside these bounds a larger reward penalty is incurred (proportional to $\rewardout$). The sum of the reward associated with each health marker defines the reward of a patient's health state. In Equation~\ref{eq:rs} we define this function explicitly for a 30 year old female patient (note that in practice the bounds are sex and age dependent)

\begin{align}
    \Rs &= \lineardropoffreward(\spotwo, 0.5, 0.25, 88\%, 95\%) \; + \nonumber\\
    &\quad\;\lineardropoffreward(\paotwo, 0.5, 0.25, 75\;\mmHg, 95\;\mmHg) \; + \nonumber\\
    &\quad\;\lineardropoffreward(\rr, 0.5, 0.25, 12\;\breathspm, 18\;\breathspm) \; + \nonumber\\
    &\quad\;\lineardropoffreward(\ieratio, 1, 4, 0.3, 0.5) \; + \nonumber\\
    &\quad\;\lineardropoffreward(\pplat, 1, 4, 0\;\cmh, 30\;\cmh) \; + \nonumber\\
    &\quad\;\lineardropoffreward(\ph, 1, 1, 7.3, 7.45) \; + \nonumber\\
    &\quad\;\lineardropoffreward(\hr, 1, 4, 74\;\breathspm, 81\;\breathspm)
    \label{eq:rs}
\end{align}

\begin{equation}
    \footnotesize
    \lineardropoffreward(x,\rewardin,\rewardout,\lowerb,\upperb) = \begin{cases}
        \rewardin & \text{if } \lowerb \leq x \leq \upperb, \\
        \rewardin - \rewardout (\lowerb - x) & \text{if } x < \lowerb, \\
        \rewardin - \rewardout (x - \upperb) & \text{if } x > \upperb.
    \end{cases}
\label{eq:rcomponent}
\end{equation}

In order, Equation~\ref{eq:rs} encodes oxygenation goals ($\spotwo$, $\paotwo$), respiratory goals ($\rr$, $\ieratio$, $\pplat$), blood chemistry goals ($\ph$), and goals pertaining to vital signs ($\hr$). A glossary for these terms is made available in Table~\ref{tab:variables}.

The {ventilator actions are evaluated} by summing the weighted magnitude of the actions within sensible lower and upper action bounds (represented as vectors $\va_{\lowerb}$ and $\va_{\upperb}$ respectively) as shown in Equation~\ref{eq:ra}. If the actions values are large within these bounds, then the reward is highly negative, and such actions will be avoided by our policies. This places a penalty on selecting aggressive, high-pressure ventilator inputs which might harm a patient. In practice, we introduce a weight vector $\vec{w}$ which allows us to control how much relative penalty certain action values impose 


\begin{align}
    \Ra &= - \sum_{i=1}^{\dimA} \; \vec{w}_i \frac{
        \va_i  - \va_{\lowerb, i}
    }{
        \va_{\upperb, i} - \va_{\lowerb, i}
    }
    \label{eq:ra}
\end{align}

\section{Policies}
\label{sec:algorithms}


\begin{figure*}[h] 
    \centering
    \begin{minipage}[b]{0.48\textwidth}
        \begin{subfigure}[b]{\linewidth}
            \centering
            \includegraphics[height=1.2cm]{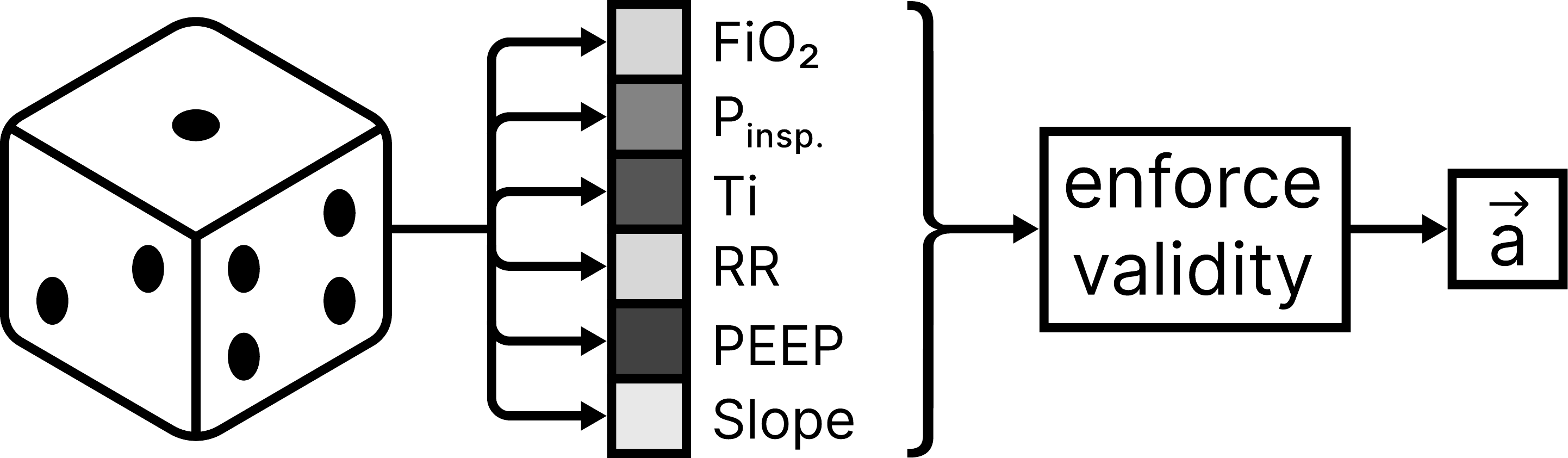}
            \caption{Actions are {randomly} selected from the space of all \textit{valid} ventilator actions (i.e. action-values are in bounds and do not conflict with one another). The patient's health state is not considered.}
            \label{subfig:random}
        \end{subfigure}
        \vspace{0.8mm}
        
        \begin{subfigure}[b]{\linewidth}
            \centering
            \includegraphics[height=1.2cm]{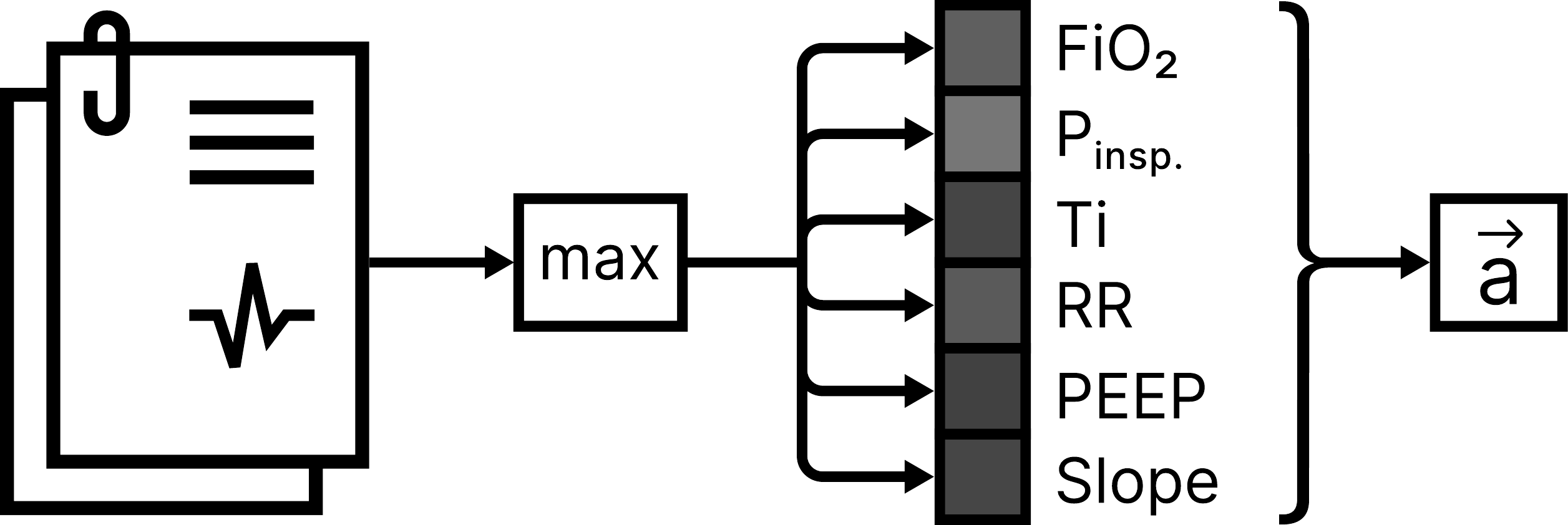}
            \caption{Actions are selected based on the {maximum allowed intervention} as recommended by clinical guidelines (the \ardsnet protocol). This method is implemented for comparison only: maximizing intervention is \textit{non-protective}, and may cause ventilator-induced lung injuries.} 
            \label{subfig:maxint}
        \end{subfigure}
        \vspace{0.8mm}
        
        \begin{subfigure}[b]{\linewidth}
            \centering
            \includegraphics[height=1.4cm]{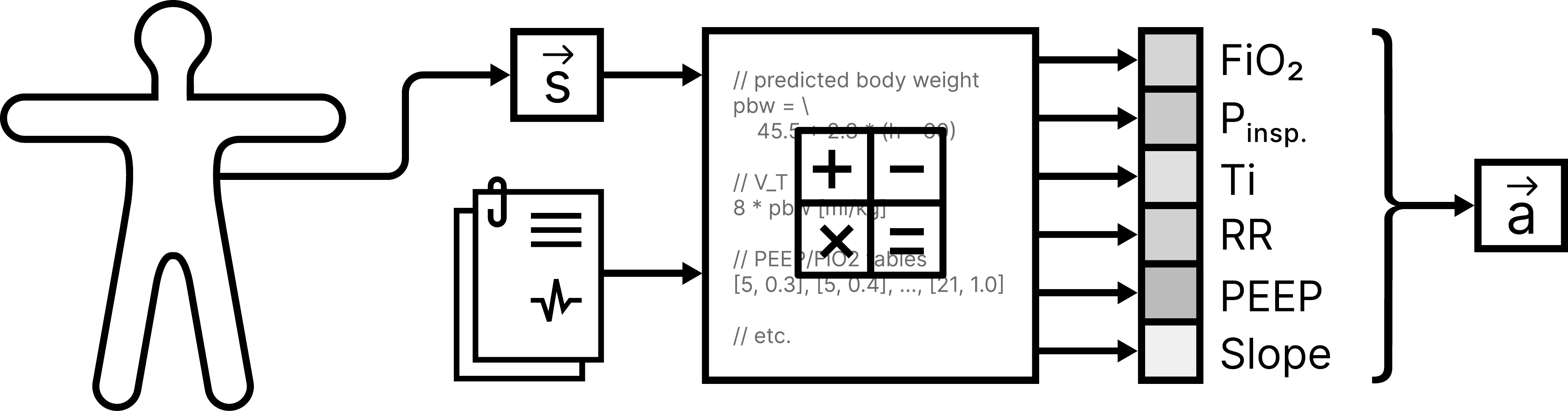}
            \caption{Actions are calculated based on the patient's health state according to the {\ardsnet protocol}; a set of clinical guidelines, recommendations, and best-practices based on years of aggregated medical evidence. The protocol accounts for a patient's current response to ventilation (e.g. oxygenation, $\ph$) and anthropometric parameters (e.g. sex, height).}
            \label{subfig:ardsnet}
        \end{subfigure}
        \vspace{0.8mm}
        
        \begin{subfigure}[b]{\linewidth}
            \centering
            \includegraphics[height=2.7cm]{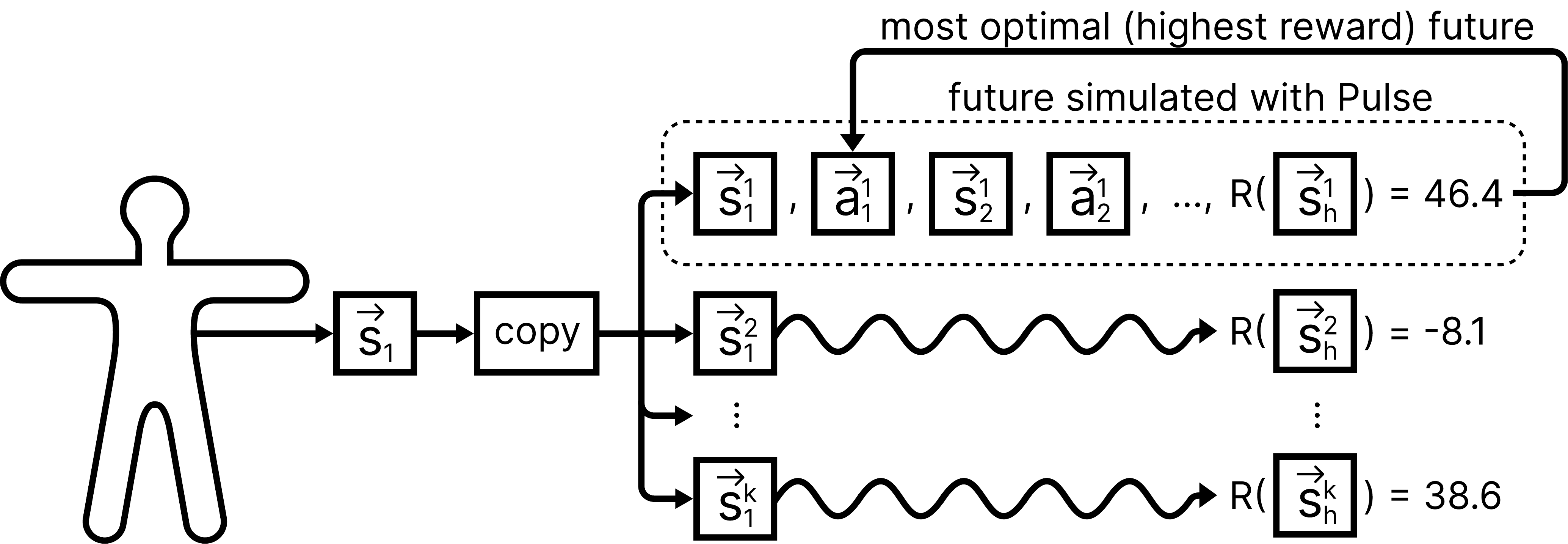}
            \caption{The patient's health state is copied $K$ times and their response to various ventilator actions is simulated using the Pulse model out to a horizon $H$. We use the $\curly$ arrow to represent a full simulation of successive states and actions, and explicitly illustrate one such future in the dotted rectangle. The patient's health in these simulated futures is evaluated with the reward function, and the immediate action which results in the optimal future (highest reward) is executed ($\vec{a}^1_1$ in this example). This method uses \textit{sampled} futures from a \textit{model} (Pulse) to \textit{predict} the optimal \textit{control} inputs (hence {Sampling-based Model Predictive Control} or {SMPC}).}
            \label{subfig:smpc}
        \end{subfigure}
    \end{minipage}
    \hfill
    \begin{minipage}[b]{0.48\textwidth}
        \begin{subfigure}[b]{\linewidth}
            \centering
            \includegraphics[height=5.4cm]{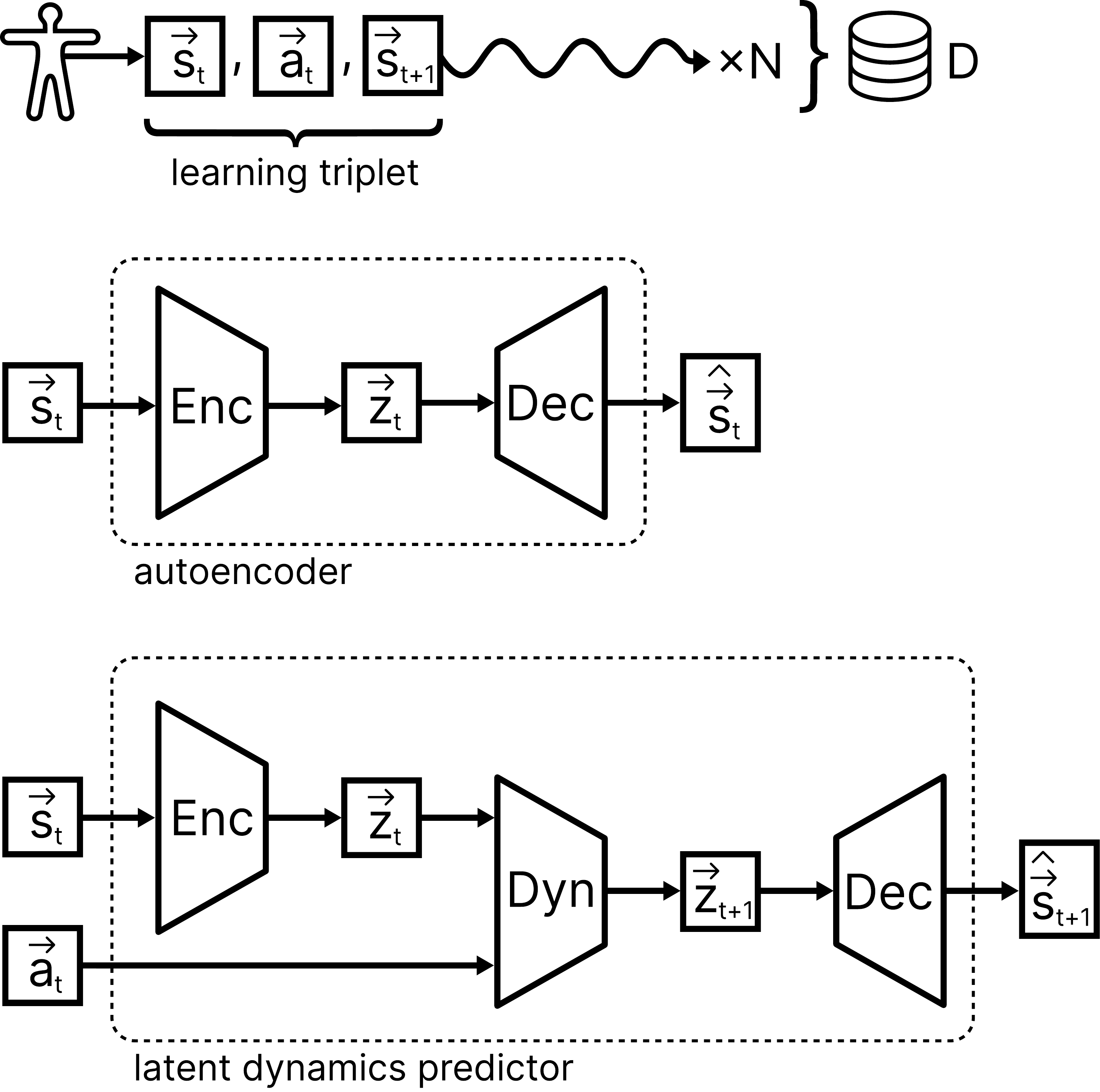}
            \caption{A dataset $D$ is first generated from many random simulations, gathering many examples of how we transition to $\vec{s}_{t+1}$ from $\vec{s}_t$ and $\vec{a}_t$. An \textit{autoencoder} neural network learns a low-dimensional \textit{embedding} of the patient's health state through a learning signal defined by reconstruction error between $\vec{s}_t$ and $\hat{\vec{s}}_t$. A second neural network learns how a patient's respiratory dynamics evolve in this low-dimensional embedding space (the {Embed To Control} paradigm) through a learning signal defined by prediction error between $\vec{s}_{t+1}$ and $\hat{\vec{s}}_{t+1}$. This learned dynamics model is used in lieu of Pulse to simulate futures and determine the best immediate action using {SMPC} identically as in Figure~\ref{subfig:smpc}. Computing respiratory dynamics with Pulse is significantly slower than \textit{estimating} these respiratory dynamics with this neural network, thus enabling techniques that leverage many samples/simulations (e.g. SMPC or MPPI).}
            \label{subfig:e2smpc}
        \end{subfigure}
        \vspace{0.1mm}
        
        \begin{subfigure}[b]{\linewidth}
            \centering
            \includegraphics[width=\linewidth]{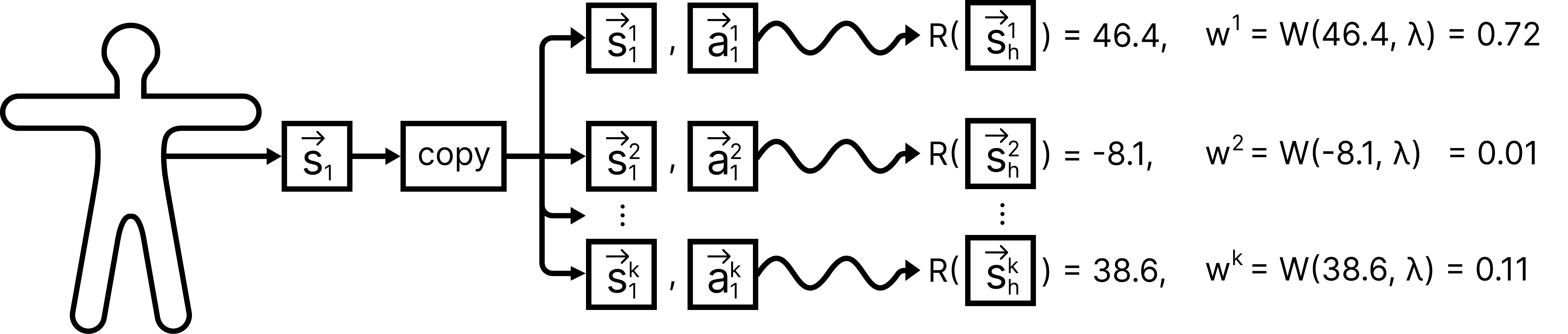}
            \caption{The {Embed To Control} paradigm is again used as in Figure~\ref{subfig:e2smpc}, but {Model Predictive Path Integral Control} ({MPPI}) is used in lieu of SMPC. In other words, the state we transition to when applying an action $\vec{a}_t$ on state $\vec{s}_t$ is computed with the learned functions rather than with Pulse like so: $\vs_{t+1} = \dyn(\vs_t, \va_t) = \dec(\dynlatent(\enc(\vec{s_t}), \vec{a_t}))$. MPPI predicts an optimal action by combining reward-weighted action sequences (and is thus more robust to disturbances), rather than simply considering the action that results in the highest reward. In the above example, the optimal immediate action would be computed as a weighted sum of the sampled action vectors: $\vec{a}^{*} = \sum_{i=1}^k w^i \vec{a}^i_1 = 0.72 \vec{a}^1_1 + 0.01 \vec{a}^2_1 + \dots + 0.11 \vec{a}^k_1$. }
            \label{subfig:e2smppi}
        \end{subfigure}
    \end{minipage}
    \caption{An overview of the policies used to control ventilators in this work: a) Random Selection, b) Maximum Intervention, c) \ardsnet Protocol, d) Sampling-based Model Predictive Control, e) Embed To Control with Sampling-based Model Predictive Control, and f) Embed To Control with Model Predictive Path Integral Control. } 
    \label{fig:algorithms}
\end{figure*}

\subsection{Random Selection}
\label{sec:rand}

We select a random ventilator action $\va$ by sampling independently from $\dimA$ uniform distributions over clinically reasonably bounds $U_{\fiotwo}(0, 1)$, $U_{\pip}(1, 30)$, $U_{\ti}(0.1, 3)$, $U_{\rr}(1, 30)$, $U_{\peep}(1, 25)$, and $U_{\textnormal{slope}}(0, 1)$ to construct the action $\va$. Physically impossible actions are made valid (e.g. $\peep$ cannot be higher than $\pip$, so it is adjusted down if such a sample is encountered). This approach is mainly used as a baseline and for generating transition data (See Section~\ref{sec:e2csmpc}).

\subsection{Maximum Intervention}

The patient's anthropometric characteristics (sex, height, etc.) are used to determine the maximum recommended intervention according to the \ardsnet protocol. This results in selecting inputs that result in the maximum allowed tidal volume $\tidalvolume$. We set $\fiotwo=1.0$ and $\peep=18$--$24$ across all patients. Maintaining these high-pressure settings is \textit{non-protective} \cite{zou2024ventstratsprotect} and we expect to see this approach make the patient enter a worse health state (i.e. lower reward).

\subsection{\ardsnet Protocol}

We compute a ventilator action by using the guidelines specified in the \ardsnet protocol. We implement this as a series of conditional rules that draw upon lookup tables established from clinical data \cite{slutsky2000mechanical, de2002ardsnet, grasso2007ardsnet}. We begin by computing the patient's predicted body weight to determine the ideal initial $\tidalvolume$. We set $\rr$ to achieve this and then adjust the magnitude of $\rr$ accordingly to achieve a $\ph$ between $7.3$--$7.45$ and $\pplat < 30 \;\cmh$. $\peep$ and $\fiotwo$ are selected according to a predefined \ardsnet table specifying valid combinations. The magnitude of the values is made proportional to the distance from a $\pao$ in the range $55$--$80 \;\mmHg$ and a $\spo$ in the range $88$--$95\%$. The steps of the \ardsnet protocol pertaining to spontaneous breathing trials are not implemented. 


\subsection{Sampling-based Model Predictive Control}

Model Predictive Control (MPC) optimizes actions over some time horizon by successively optimizing over immediate shorter time horizons. MPC generally solves the shorter time horizon problems analytically by taking advantage of an explicit model of the system dynamics (i.e. the respiratory dynamics modeled by Pulse). An analytical approach is not feasible in this case due to the complexity of the system dynamics, so we determine the optimal sequence of actions using sampling (hence Sampling-based MPC, or SMPC). The sample with the highest reward is associated with the optimal sequence of ventilator actions, and we execute the first action from said optimal sequence (see Figure~\ref{subfig:smpc}). 


The success of SMPC is largely based on 1) the number of samples $K$ used, 2) the horizon $H$ that we simulate out to, and 3) the approach to generating potential series of actions. In this work, we use $K=\Kusedsmpc$ and $H=\Hused$, meaning that $\Kusedsmpc$ possible sequences of ventilator actions are considered $H\Delta t=\Hused\times\deltat=120$ minutes $=2$ hours into the future. Actions can be generated using any practical method that adequately explores the space of all valid ventilator actions (e.g. uniform random sampling, maximizing space filling metrics over the action space, stratified sampling \cite{kochenderfer2019algorithms}, etc.). In this case we use uniform random sampling as described in Section~\ref{sec:rand}.



This approach has two key shortcomings: 1) computing updates with Pulse is relatively slow, and 2) in a real-world setting the patient-ventilator dynamics from Pulse may not fully capture the patient-specific dynamics or other disturbances. In other words, it may be possible to more accurately and rapidly model the dynamics by learning from data produced by a specific patient-ventilator system.

\subsection{Embed To Control with Sampling-based MPC}
\label{sec:e2csmpc}

We use the Embed To Control (E2C) paradigm to address the previously discussed shortcomings of SMPC. E2C works by 1) learning a compact, lower dimension, latent representation of the patient's health state using an autoencoder, and 2) learning how the patient's health state evolves given some ventilator action using a predictive dynamics model that operates \textit{in the latent space of the autoencoder}.


To generate training and validation data we simulate the full benchmark ($\numpat$ patients) {twice}, using the random policy (Section~\ref{sec:rand}) for $200$ total runs. Each of the $200$ runs has $96$ consecutive actions spaced at $\Delta t =\deltat$ minutes, producing $(96-1)\times 200 = 19000$ instances of a state-action-state triplet $(\vs_t, \va_t, \va_{s+1})$. Each of these instances represents an example of how a patient's health state evolves given a specific ventilator action.  

The autoencoder's encoder ($\enc$) is implemented as a two-layer neural network with input, hidden, and output sizes $\dimS, 16, 6$ and with rectified linear unit (ReLU) activations after each layer. The decoder ($\dec$) is implemented identically but with layer sizes reversed: $6, 16, \dimS$. States $\vs$ are encoded into latent representations $\vz$ and then decoded, producing an estimate of the original state $\hat{\vs}$. 

The \textit{latent} dynamics model ($\dynlatent$) is implemented as a two-layer neural network with input, hidden, and output sizes $12, 8, 6$ and with rectified linear unit (ReLU) activations after each layer. Inputs are a concatenated vector $(\enc(\vs_t), \va_t)$ of size $6+6=12$, and outputs are a predicted \textit{next latent state vector} $\hat{\vz}_{t+1}$ of size $6$. By using the previously trained decoder we can estimate the next state given the current state and action as in Equation~\ref{eq:learnrollout} 

\begin{equation}
    \hat{\vs}_{t+1} = \dec(\dynlatent(\enc(\vec{s_t}), \vec{a_t}))
    \label{eq:learnrollout}
\end{equation}

The training process occurs in two stages: the autoencoder is first trained, its weights are frozen, and the latent dynamics model is then trained (as illustrated in Figure~\ref{subfig:e2smpc}). Both neural networks are optimized using Adam \cite{kingma2014adam} for $256$ epochs each, with batch sizes of $64$.



The true (Pulse) patient-ventilator dynamics can now be rapidly estimated using Equation~\ref{eq:learnrollout}. Thus, SMPC can be executed in a manner identical to before---except the outcomes of candidate action sequences are now estimated more rapidly using the learned model, allowing us to use considerably more samples ($K=1024$) and thus explore the space of potential actions more thoroughly.

\subsection{Embed To Control with MPPI}

With SMPC, when multiple high-reward action sequences are discovered, only the highest reward sequence is executed. This fails to take advantage of the exploration of the state space that has \textit{already been conducted}. With Model Predictive Path Integral Control (MPPI) we introduce a hyperparameter $\lambda$ which controls how \textit{all} action sequences are combined into an approximately optimal action sequence. Put plainly, MPPI creates a reward-weighted combination of \textit{all sampled action sequences}, and the sharpness of the weights distribution is controlled by $\lambda$ \cite{williams2017model}.

We illustrate a simple version of MPPI in Figure~\ref{subfig:e2smppi}. Note that SMPC is equivalent to MPPI when $\lambda=\infty$, i.e., all weights are zero apart from the one associated with the highest reward trajectory (the strategy then becomes `pick the actions that produce the highest reward'). For our MPPI approach, we again use the E2C paradigm and leverage learned latent dynamics to rapidly simulate the outcomes of potential action sequences using Equation~\ref{eq:learnrollout} ($K=1024$).



\section{Results}

\subsection{Optimal Control Methods Outperform Baseline Methods}

\begin{figure}[h]
    
    \begin{subfigure}{0.45\textwidth}
        \centering
        \includegraphics[width=\textwidth]{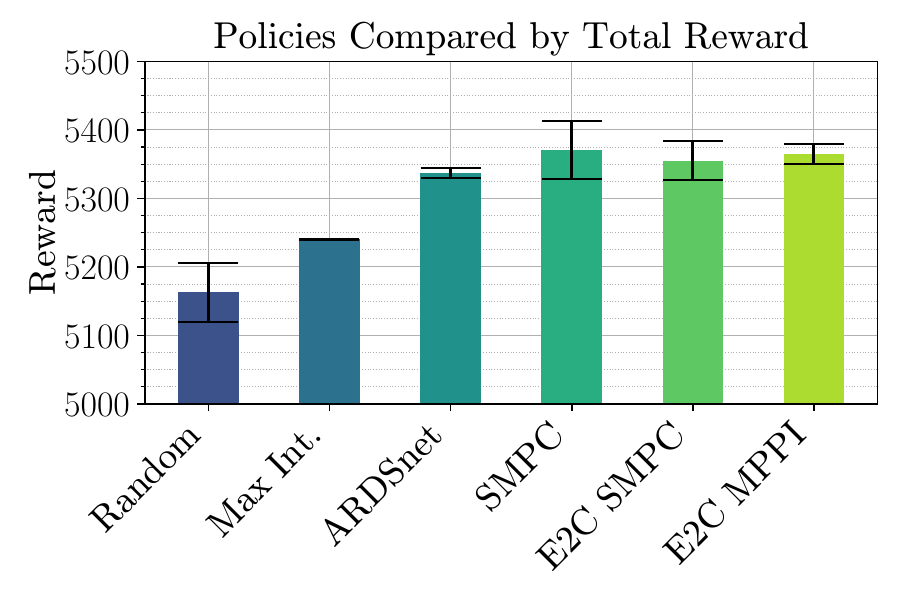}
        \vspace{-7mm}
    \end{subfigure}
    \caption{The total accumulated reward that each policy has earned at the end of the task, averaged over all patients ($N=\numpat$). Shown are $95\%$ confidence intervals. } 
    \label{fig:rewards}
\end{figure}


We begin by {comparing each policy in terms of total accumulated reward}, as shown in Figure~\ref{fig:rewards}. As expected, the random policy scores poorly as it leverages no information regarding the patient's health state when computing actions. The maximum intervention policy also scores poorly as it is penalized for selecting needlessly aggressive (i.e. sustained high-pressure) ventilator actions, thus validating our reward function formulation.

The SMPC policy slightly outperforms the \ardsnet policy. This is likely due to SMPC's ability to sample arbitrary continuous actions in $\realR^\dimA$, whereas the \ardsnet policy is confined to a limited set of specific combinations of action values (e.g. $\fiotwo$ and $\peep$ must be chosen according to a discrete table of options), thus enabling less precision. In reality, limiting the selection of certain action values has additional utility in a medical care setting: such protocols can be easily remembered, interpreted, and mentally-computed by care providers, unlike policies based on optimal control. 

Finally, we note that our MPPI policy---and our policies leveraging the E2C paradigm---accumulate comparable reward to \ardsnet throughout the task, implying that such techniques offer a promising approach to automatic ventilator management for patients with ARDS. We note that these methods discover useful ventilation strategies despite only having access to expert knowledge encoded in the reward function. In other words, {these methods learn effective ventilation management strategies \textit{without} access to explicit ventilator management procedures or guidelines} (such as those defined in the \ardsnet protocol). 


This outlines a unique advantage of these techniques; certain health markers can be weighted as more important than others with simple adjustments to the reward function, and ventilator management strategies that focus on optimizing these markers will then be automatically discovered. For example, a higher penalty could be placed on ventilator actions that result in worse oxygen saturation, or extremely high heart rates, thus allowing for more flexibility than strict rules-based procedures.

\begin{table}[h]
  \centering
    \begin{tabular}{@{}l p{3mm}p{3mm}p{3mm}p{3mm}p{3mm}p{13mm}p{3mm}p{3mm}@{}} 
        \toprule
    & $\tidalvolume$ & RR & $\spotwo$ & $\ieratio$ & HR & BP (sy./di.) & $\pplat$ & pH \\
    & (\%)  & (\%)  & (\%)  & (\%)  & (\%)  & (\%)  & (\%) & (\%) \\
    \midrule
    Random      & 33 & 97  & 93 & 79  & 55 & 100/77 & 100 & 19 \\
    Max Int.    & 12  & 100 & 99 & 100 & 86 & 100/34 & 100 & 15 \\
    \ardsnet     & 89 & 100 & 99 & 98  & 82 & 100/85 & 100 & 89 \\
    SMPC       & 87 & 99  & 98 & 96  & 83 & 100/83 & 100 & 86 \\
    E2C SMPC    & 84 & 97  & 98 & 95  & 79 & 100/84 & 100 & 84 \\
    E2C MPPI    & 86 & 99 & 98 & 95  & 78 & 100/85 & 100 & 87 \\
    \bottomrule
  \end{tabular}
  
  \caption{The percentages of the ventilation period where the patient's health markers stayed within a healthy range for their sex and age (averaged and rounded to the nearest percentage over all patients). Higher percentages mean that a given strategy optimized that patient's health during ventilation with respect to that specific marker. We note that the systolic blood pressure and plateau pressure always stayed within healthy ranges, and that respiration rate and oxygen saturation stayed within healthy ranges during over $90\%$ of the ventilation period \textit{across all methods}. Abbreviations: tidal volume $\tidalvolume$, respiration rate $\rr$, oxygen saturation SpO$_2$, inspiratory expiratory ratio $\ieratio$, heart rate $\hr$, blood pressure BP (systolic, diastolic), plateau pressure $\pplat$ \& potential of hydrogen pH.}
  \label{tab:ranges}
\end{table}


We further {compare each policy in terms of the patients' health markers}. In Table~\ref{tab:ranges}, we compute the proportion of time that various health markers were within healthy ranges (averaged over all patients). We note that the random and maximum intervention approaches fail as effective ventilator management strategies, as important ARDS improvement markers are not in-range for considerable portions of the simulation. 


Moreover, all optimal control policies perform similarly in terms of achieving in-range health markers, with heart rate, pH, and $\tidalvolume$ showing the most separation between the optimal control policies. Finally, we note that the baseline \ardsnet slightly outperforms or performs comparably to all optimal control policies, particularly with respect to $\tidalvolume$ and $\ph$, both of which are explicitly specified as targets in the \ardsnet protocol. 


The deviations between a policy's total accumulated reward and health markers performance suggest a misalignment between the goals encoded in the reward function and the desired health goals for ARDS patients. In the optimal control setting, if a patient-specific health goal is desired, we can adjust the policies by updating the reward function appropriately.




\subsection{Neural Networks Model Dynamics Effectively \& Efficiently}

The SMPC policy performs nearly identically when using the E2C paradigm and when using Pulse to simulate the patient's respiratory dynamics (Figure~\ref{fig:rewards}). This implies that {our neural network latent dynamics model is effective in its approximation of Pulse's respiratory dynamics}, and has indeed learned general respiratory dynamics from data. 

{The neural network models are also time efficient}, in that they can rapidly approximate dynamics. Specifically, it takes $156$ seconds to exactly simulate a single timestep with Pulse, but approximately $2$ seconds to approximate them using E2C models. Faster dynamics computations enables more samples to be computed in the same amount of time, improving our approximations of the optimal ventilator action, and thus our performance on the task as a whole.

We currently learn general respiratory dynamics across data produced from multiple patients, but patient-specific dynamics could be learned by using the data from a single patient, enabling more personalized healthcare. We leave this investigation to future work.



\section{Conclusions}

In this work we have introduced a repeatable benchmark for testing ventilator strategies for Acute Respiratory Distress Syndrome (ARDS) patients (based on the Pulse Physiology Engine). We implement and compare policies based on clinical recommendations (the \ardsnet protocol), optimal control theory, and learning-based techniques (representing respiratory dynamics with neural networks). 

We find that optimal control policies in conjunction with learning-based techniques can discover effective ventilator management strategies `from scratch'---without access to explicit ventilator management procedures or guidelines (such as those defined in the ARDSnet protocol), thus offering a promising and automated approach to ventilator management for patients with ARDS.

Avenues for future work include:

\begin{itemize}
    \item Extending our techniques to other respiratory conditions, and incorporating safety constraints into the optimal control of ventilators.
    
    \item Learning to encode histories of states and actions (rather than only states) and leveraging more powerful sequence-to-sequence learning models (e.g. Transformers).

    \item Validating that our approaches trained on simulated data generalize to real-world patient-ventilator data.
\end{itemize}

\appendix

\subsection{Reproducibility}

All experiments were executed on a workstation with an AMD Ryzen Threadripper 3970X 32-Core 64-Thread Processor (3.69 GHz) CPU and 128 GB of usable RAM. The training of neural networks was executed on NVIDIA GeForce RTX 3090 GPUs. The code and data required to reproduce these results is available at \href{https://github.com/sisl/ventilators}{https://github.com/sisl/ventilators}.

\subsection{Glossary}

All variables discussed in this work are defined in Table~\ref{tab:variables}.

\begin{table*}[htbp]
    \centering
    \begin{tabular}{@{}p{4.5cm} p{1.5cm} p{8cm}@{}}
        \toprule
        \textbf{Variable} & \textbf{Units} & \textbf{Description} \\
        \midrule
        Fraction of inspired oxygen ($\fiotwo$) & -- & The ratio of oxygen in the inspired air that is involved in alveolar gaseous exchange, as compared to the total amount. \\
        Inspiratory pressure ($\pip$) & $\cmh$ & Pressure during inspiration. \\
        Inspiratory period ($\ti$) & s & Duration of inspiration. \\
        Respiratory rate ($\rr$) & breaths/min & Number of breaths per minute. \\
        Positive end expiratory pressure ($\peep$) & $\cmh$ & Pressure in airways at end of expiration. \\
        Slope & -- & Controls rate of airway pressure increase. \\
        \midrule
        ECG III & mV & Electrocardiogram. Electrical difference between left arm and left leg. \\
        PLETH & -- & Photoplethysmography. A measure of the pulse oximeter signal, representing changes in arterial blood flow. \\
        $\cotwo$ pressure & mmHg & Pressure of carbon dioxide dissolved in arterial blood. \\
        Heart rate & beats/min & Heart rate. \\
        Diastolic blood pressure & mmHg & Pressure in arteries during relaxation of heart. \\
        Systolic blood pressure & mmHg & Pressure in arteries during contraction of heart. \\
        $\spotwo$ & \% & Oxygen saturation. \\
        $\etcotwo$ & \% & End-tidal carbon dioxide fraction. \\
        awRR & breaths/min & Airway respiratory rate. \\
        T & \textdegree C & Core temperature. \\
        $\paotwo$ & mmHg & Arterial oxygen pressure. \\
        Inspiratory flow & L/min & Flow rate during inspiration. \\
        Expiratory flow & L/min & Flow rate during expiration. \\
        \ieratio & -- & Ratio of inspiratory to expiratory time. \\
        Total lung volume & L & Maximum volume of air lungs can hold. \\
        Peak airway pressure (Paw) & $\cmh$ & Highest pressure in airways. \\
        Plateau pressure ($\pplat$) & $\cmh$ & End-inspiratory airway pressure (after airflow has ceased). \\
        Tidal volume ($\tidalvolume$) & mL & Volume of air moved in/out of lungs per breath. \\
        Minute ventilation & L/min & Volume of gas inhaled/exhaled per minute. \\ 
        Mean airway pressure  & $\cmh$ & Average pressure in airways. \\ 
        Dynamic lung compliance  & L/$\cmh$ & Measurement of lung compliance. \\ 
        Static pulmonary compliance & -- & Lung compliance during no gas flow. \\
        Dynamic pulmonary compliance & -- & Lung compliance during gas flow. \\
        pH & -- & Measurement of the acidity or alkalinity of blood. \\
        \bottomrule
    \end{tabular}
    \caption{The action and state variables involved in this work. `--' indicates that the measurement is unitless. Note that some state variables are read from both the simulated ventilator and from physiological sensors (i.e. they are two dimensional). }
    \label{tab:variables}
\end{table*}


\printbibliography



\end{document}